\documentstyle[11pt,newpasp,twoside,epsf]{article}
\def\mpe{M_{\rm pe}}

\def\edcomment#1{\iffalse\marginpar{\raggedright\sl#1\/}\else\relax\fi}
\reversemarginpar

\begin{document}
\title{Mass Loss from Globular Clusters}
\author{Douglas C. Heggie}
\affil{University of Edinburgh, Department of Mathematics and
 Statistics, King's Buildings, Edinburgh EH9 3JZ, UK}

\begin{abstract}We consider both observational and theoretical issues
 related to dynamical mass loss from the old globular star clusters of
the Galaxy.  On the
observational side the evidence includes tidal tails and extratidal
extensions, kinematic effects, effects on the mass function, and
influences on the statistical properties of surviving objects.
Even for isolated clusters, the theoretical issues are not fully
understood.  The effects of a steady tide (i.e. for a cluster in a
circular orbit) include the imposition of a tidal boundary, and lowering of
the escape energy.  Less familiar, however, are the effects of {\sl
induced} mass loss.  Even the definition of an ``escaper'' is not
straightforward.  When mass loss is driven by relaxation, as in $N$-body
models, the rate of loss of mass does not scale in a simple way with
the relaxation time.  Reasons for this include the very long time
scales on which stars escape even with energies above the escape
threshold.  For the realistic case of unsteady tides it is still
unclear under what circumstances mass loss is dominated by relaxation
or tidal heating.
\end{abstract}

\section{Introduction}

Among the reasons for studying this problem are the following:

\begin{enumerate}
\item There is a growing amount of direct observational evidence showing that
globular clusters lose mass (cf. Sec.2 and the paper by Meylan in
these Proceedings, and several of the poster papers).
\item Preferential loss of stars of low mass causes the mass function
to evolve from its primordial form (cf. Sec.2).
\item It affects the statistical distribution of cluster properties in
the galaxy, causing the death of clusters of low mass or large radius (Sec.2).
\item It has much in common with the problem of disruption of
satellite galaxies, in which there has been much interest recently
(e.g. Kroupa 1997, Ibata et al. 1997).
\item Decayed globular clusters may be a substantial
contribution to the stellar halo of the Galaxy (cf. Ashman \& Zepf 1998).
\item On the theoretical side, understanding mass loss turns out to be
a vital issue in establishing the right boundary conditions to use in
simplified models, such as Fokker-Planck models (Takahashi \&
Portegies Zwart 1998, 1999; Spurzem \& Takahashi 2000).
\item Also on theory, mass loss turns out to be a key to understanding
how to scale $N$-body models with $N$ (cf. Sec.3).
\end{enumerate}

The emphasis in this review is on theoretical problems, but it begins
with a summary of the observational position.

\section{Observational Issues}

\subsection{Tidal tails and extensions}

Two surveys (Grillmair et al. 1995; Meylan et al. 1999, Leon, Meylan
\& Combes 2000, and Meylan,
these Proceedings) and some data on individual objects (Zaggia et
al. 1997 for M55, Testa et al. 2000 for M92) have shown the presence of
stars beyond the conventional tidal radius.  In the case of the
surveys the shapes tend to resemble those of tidal tails observed in
simulations (cf. Fig.1), while the studies of M55 and M92 revealed
roughly spherical extensions.

\begin{figure}
\plotfiddle{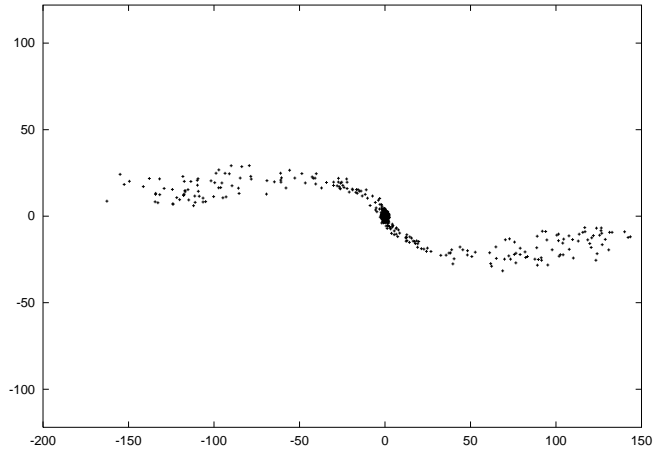}{2.5 in}{-90}{35}{35}{-150}{190}
\caption{Tidal extensions in an $N$-body model.  Here the tide is
steady, caused by the host galaxy far away along the vertical axis, and
mass loss is caused by relaxation, and
``induced'' mass loss (Sec.3).  As with most simulations, however,
near the cluster the tidal extension is in the direction of the
perturbing field, while further away it is parallel to the direction
of orbital motion.  Initial model:  King, with $W_0 = 3$, $N=10^4$; time: 1
galactic orbital period; units: $N$-body units.}
\end{figure}

These studies raise a number of theoretical questions.  In Grillmair
et al. the cluster with the strongest tidal tail is M2 (=NGC7089), and
modelling based on this data (Johnston et al. 1998) suggests that its lifetime is of
order 1-3Gyr.  Theoretical estimates, however, imply that this object
is losing mass at one of the very slowest rates of all Galactic
globular clusters, on a time of order 10--20 Hubble times (Gnedin \&
Ostriker 1997, Aguilar et al. 1988).  The direction of the tail is
not, at first sight, consistent with the most plausible cause
(shocking during a recent passage through the Galactic disk).  Meylan
et al. themselves draw attention to two clusters in their sample with
very similar orbits (and hence strength of tidal effects) but tails of
very different strength.  Furthermore the dependence on radius is
flatter than in any of their simulations.  The presence of
a nearly {\sl spherical} extra-tidal extension, as in the case of M92,
raises the question of what the real value of the tidal radius is.

\subsection{Kinematic Evidence}

By studying the radial velocities of large numbers of stars in M15,
Drukier et al (1998) found a slight rise in the velocity dispersion at
large radii (but still well within the tidal radius).  This was
interpreted at
first as evidence of tidal heating, but it was then realised, using
$N$-body modelling, that this effect can occur even in clusters on
{\sl circular} galactic orbits.  (This can also be seen in Giersz \&
Heggie 1997.)  In this case the tide is steady (in a rotating frame),
and its effect cannot be described as ``heating'' (Sec.3).

In two individual clusters, stars are observed with {\sl radial}
velocity exceeding the escape velocity (Gunn \& Griffin 1979; Meylan,
Dubath \& Mayor 1991).  As with extratidal extensions, the
significance of these observations depends on the reliability of the
models which yield the value of the escape speed.  If taken at face
value, however, such observations are usually interpreted as ejecta
from three-body interactions (e.g. Davies 1992).  

An alternative interpretation is suggested by recent $N$-body
simulations (Fig.2).  They show that the number of stars inside the
tidal radius with energies above the energy of escape (``potential
escapers'') is surprisingly large.  Furthermore, the maximum mass of
this population, $\mpe$, decreases surprisingly slowly with increasing
$N$: $\mpe/M(0) \sim 0.1 (N/3\times10^4)^{-0.23}$, where $M(0)$ is the
initial cluster mass.  Therefore even for a real cluster the
proportion could be as large as $\sim$5\%, and a considerably larger
proportion of the {\sl current} cluster mass.  These results apply to
models on circular galactic orbits, however.

\begin{figure}
\plotfiddle{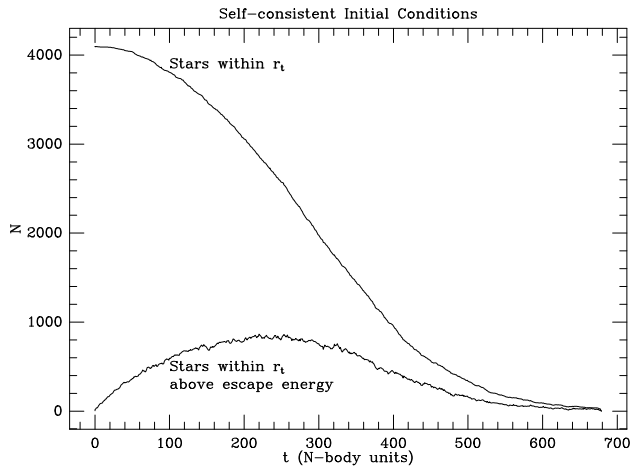}{2.5 in}{-90}{35}{35}{-150}{190}
\caption{The total cluster mass (upper) and the mass of the ``potential
escapers'' (lower; see text) in a 4K-model.  The initial and other conditions are
those of the collaborative experiment described in Heggie et
al. 1998, except that initial conditions are generated as in Heggie \&
Ramamani 1995.}
\end{figure}

\subsection{Evidence from Mass Functions}

At one time there seemed very clear evidence that the slope of the
mass function in observed clusters is correlated with galactocentric
radius and $z$, consistent with an interpretation in terms of preferential loss
of low-mass stars (Capaccioli et al. 1993).  The best that can be said now (Piotto \& Zoccali
1999) is that, while there seems to be a trend, assessing its
statistical significance is hard.

One cluster in which the effect of mass loss may be most pronounced is
NGC 6712, which definitely appears to have a flatter mass function
than any other comparable cluster (de Marchi et
al. 1999), and has received considerable attention.  Takahashi \&
Portegies Zwart (1999) gave an example of an $N$-body model which,
after losing 99\% of its mass, resembles this cluster.  This is not
implausible for a cluster whose present mass is estimated at
$10^5M_{\sun}$, and  a
tentative reconstruction of its orbit (Dauphole et al. 1996) suggests
a very small galactocentric radius and therefore heavy tidal mass loss.  A large
mass loss rate for this object was also estimated by Gnedin \&
Ostriker (1997), though there is nothing about the appearance of this
cluster to suggest that it is close to death.

\subsection{Evolution of the Galactic Globular Cluster system}

There have been many studies, especially in recent years, of the
effects of mass loss and other dynamical processes on the statistical
properties of the system of Galactic globular clusters.  A few
(e.g. Aguilar et al. 1988, Gnedin \& Ostriker 1997) have
given estimates for all individual clusters about which enough is
known.  In particular, Gnedin \& Ostriker (following Fall \& Rees
1977) plot a survival diagram of
half-mass radius and cluster mass to show that the disruptive effects
of relaxation, tidal shocking and dynamical friction roughly account
for the present-day distribution of the Galactic globular clusters.
Even at a statistical level the agreement is imperfect -- there are
too many low-mass clusters -- which suggests
that either initial conditions also have a role, or the dynamical
processes are not yet sufficiently understood quantitatively.

\section{Theoretical Issues}

Several processes which lead to mass loss from star clusters have been
suggested.  In roughly the historical order in which they have
come to prominence, they are

(i) two-body relaxation, which was the main process considered until
the 1960s;

(ii) tidal effects;

(iii) mass loss from stellar evolution, which began to be allowed for
routinely in this subject only after Applegate (1986); and

(iv) massive dark halo objects, a relatively speculative mechanism
about which no more will be said here.

\subsection{Isolated Systems and Systems with a Tidal Cutoff}

Though highly idealised, the simplest problem of all would seem to be
mass loss from equal-mass systems with no external field.  In fact
this problem has been more-or-less abandoned without being completely
understood.  The essential issue is whether the time scale of mass
loss follows the relaxation time scale $t_r\sim t_{cr}N/\ln N$ or the
``strong encounter'' time scale $t^\ast_r\sim t_{cr}N$, where $t_{cr}$
is the crossing time.  It is not inconsistent to suppose that internal
processes like core collapse proceed on the time scale $t_r$ while the
time scale of escape, which may depend on energetic encounters, is
more like $t^\ast_r$.  Long ago H\'enon (1960, 1969) gave formulae for
the escape rate involving essentially $t^\ast_r$, and this prediction
is confirmed by $N$-body results (Giersz \& Heggie 1994).  Only
slightly more recently, however, Spitzer \& Shapiro (1972) gave a theory for isolated
systems which gives escape on a time scale $\propto t_r$.

These issues are not irrelevant when we turn to tidally limited
systems, at least if the tide is represented as a {\sl cutoff}.  There
is no reason why H\'enon's beautiful but little-known general formula for the escape
rate (H\'enon 1960), i.e.
\[
\dot N = - \frac{256}{3}\sqrt{2}\pi^4G^2m^2\int_{0}^{r_{max}}
r^2 dr\int\int\frac{f(E)f(E^\prime)(E+E^\prime-\phi)^{3/2}}{E^2}dE
dE^\prime,
\]
which leads to an escape time scale $\propto t^\ast_r$, is
inapplicable, and yet virtually all theorists would adopt a time scale
based on the usual relaxation time $t_r$.  One reason for this is that
so much work is based on the Fokker-Planck equation, which has no
other time scale.  Among theorists one exception is Wielen
(e.g. 1988), who argues on the basis of $N$-body data that strong
encounters greatly dominate for clusters with a few hundred stars.
Since the Coulomb logarithm varies so slowly with $N$ he concludes that
they still dominate for systems of the size of globular clusters.

\subsection{Steady Tides}

Even in a steady tide, e.g. a cluster in a circular galactic orbit, it
is useful to distinguish three effects: (i) the finite boundary,
usually taken at the radius of the Lagrangian point (cf. Spitzer
1987); (ii) the reduced escape energy; and (iii) ``induced'' mass
loss, which is described further below.

\subsubsection{Definition of Escape}  Several escape criteria are in
common use, and all are wrong.  Unless the potential of the cluster
changes, a star may have an energy above the escape energy and yet
remain within the tidal radius, $r_t$ indefinitely (Fukushige \&
Heggie 2000 and references therein).  Within the usual tidal
approximation it is possible for a star to recede to an arbitrary
distance and eventually return to the cluster (Ross, Mennim \& Heggie
1997), and so a simple distance or apocentre criterion fails.  Ross et
al. give a rigorous criterion, but in practice what is usually done in
$N$-body simulations is to measure the mass within $r_t$, and then the
escape criterion for simplified models (gas or Fokker-Planck), though
based on physical ideas, is adjusted for reasonable agreement
(e.g. Portegies Zwart \& Takahashi 1999).

\subsubsection{Induced Mass Loss}  This non-standard term is used here
to describe the following fact.  If a cluster in a steady tide loses
mass for any reason (e.g. mass loss of stellar evolution), then the
tidal radius decreases and the escape threshold changes, so that still
more mass is lost as the cluster readjusts.  This additional mass loss
is referred to as ``induced mass loss''.  If the cluster has little
mass near the tidal radius or near the escape energy then induced mass
loss is small.  In King models it is a few percent of the initial mass
loss.  In Woolley models, however, in which the
velocity distribution is a Maxwellian {\sl truncated} at the escape speed,
it is a much large factor, and in fact models with a Woolley
concentration parameter $W_0\la4$ are unstable to induced mass loss. 

Consider a cluster in which the tide is treated as a cutoff, the
cluster loses mass by stellar evolution, and relaxation is very slow.
(Weinberg (1993) gives results of such models.)  In this situation even
an initial King-like distribution evolves to one more like a Woolley
model.  It seems likely that the above instability is responsible for
the rapid loss of mass which such models appear to exhibit (Chernoff
\& Weinberg 1990, Fukushige \& Heggie 1995) at the end of their life.
Clusters in which mass loss is dominated by relaxation do not show
this behaviour.  The distinction between these two kinds of evolution
is sometimes referred to as the ``ski-jump'' problem (Fig.3), and was
highlighted by Portegies Zwart et al. (1998).

\begin{figure}
\plotfiddle{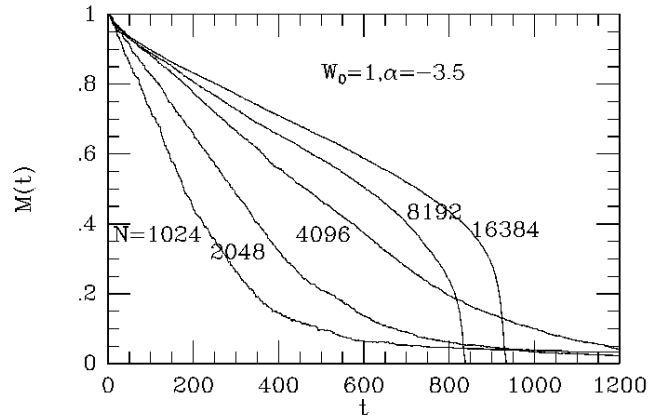}{2.5 in}{0}{35}{35}{-100}{-50}
\caption{The ``ski-jump'' problem:  mass against time for models
dominated by mass-loss through stellar evolution (large $N$), and
those dominated by relaxation (small $N$), from Fukushige \& Heggie
(1995), where all details are given.  Imagine a skier descending along each
curve!}
\end{figure}

\subsubsection{The Rate of Mass Loss}  The traditional picture (in the
absence of mass loss by stellar evolution) is that stars gain
sufficient energy to escape on the relaxation time scale, $t_r$, while escape
itself takes place on a crossing time scale.  As the latter is much
shorter when $N$ is large, it follows that the time scale of mass loss
is proportional to $t_r$.  $N$-body simulations give results that are
sufficiently different from this prediction (Fig.4) that extrapolation to, say,
$N\sim10^6$, which would be required for scaling these models to a
typical globular cluster, would yield results differing by a factor of
order two.

\begin{figure}
\plotfiddle{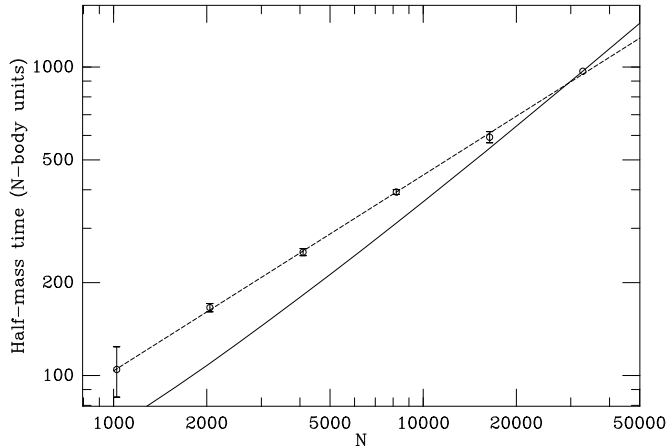}{2.5 in}{-90}{35}{35}{-150}{190}
\caption{The time scale on which half the initial mass is lost, as a
function of $N$, the initial particle number.  Initial conditions are
specified as in the collaborative experiment (Heggie et al. 1998).
The solid and dashed lines are, respectively, the result obtained by
assuming a time scale proportional to $t_r$ (fitted to the last point), and an empirical linear
fit on this log-log graph (which corresponds to a half-mass time
$\propto N^{0.63}$).}
\end{figure}

Much work has been done on this so-called ``scaling'' problem since
its discovery.  The problem is reduced somewhat if self-consistent
initial conditions are used, i.e. a model which is in equilibrium
within the tidal field, unlike a standard King model (cf. Fukushige \&
Heggie 2000).  The main problem, however, is the time scale on which
stars escape, and the resulting population of ``potential escapers''
(Fig.2).  It has been known for a long time (King, pers. comm.)  that
stars with energy above the escape energy may remain in a cluster for
many crossing times before, if ever, finding a way out, but it is only
recently that this aspect has been quantified (Fukushige \& Heggie
2000) as a function of energy.  These authors also determined the fraction of
stars with energy above the escape energy which would {\sl never} escape (if
the potential remained fixed).  How this new understanding resolves
the scaling problem is a story which has been taken up by Baumgardt
(these proceedings, and Baumgardt 2000).

\subsection{Unsteady Tides: Bulge Shocking} 

Unsteady tides usually considered are of two kinds: (i) bulge
shocking, (ii) disk shocking.  The dynamical problems they pose are
similar, and the remainder of this review concentrates heavily on
bulge shocking.  It happens only to clusters on elliptic orbits,
however, whereas disk shocking would affect any cluster except one
lying strictly in the galactic plane.

\subsubsection{Theory of Bulge Shocking}

This mechanism is often referred to as bulge {\sl heating}, and indeed
considerable emphasis is placed on its effect on the energy of stars
and clusters.  The effect of the heating, however, is to accelerate
the expansion of the halo of a cluster across the tidal boundary, and
so this mechanism is a legitimate subject of interest in this review.

Spitzer (1987) treated the problem impulsively, with qualitative
discussion of the so-called {\sl adiabatic correction} (for the fact
that the orbital period of a star within the cluster need not be much
greater than the time scale of variation in the tide).  The argument
was quantified and elaborated by Aguilar, Hut \& Ostriker (1988),
using an exponential form of adiabatic correction due to Spitzer in
another context.  The next interesting development was Weinberg's
realisation (Weinberg 1994a,b,c) that adiabatic invariants do not
protect resonant combinations of frequencies in multi-dimensional
systems (such as three-dimensional motion in a fixed cluster
potential).  (Some of the dynamical issues are ably expounded also in
Henrard 1982 and Sridhar \& Touma 1996.)  The effect of this is a
significant change from an exponential adiabatic correction to a power
law, as has been checked (with some empirical improvements) in
simulations (Gnedin \& Ostriker 1999).

One application of these results is their incorporation into
Fokker-Planck models.  This has been accomplished by Weinberg (1994c)
and by Gnedin \& Ostriker (1997), and applied to the evolution of
clusters by Murali \& Weinberg (1997) and Gnedin, Lee \& Ostriker
(1999).  Some results must be treated with caution, however, as the
formulation of Murali \& Weinberg results in bulge heating even in the case of
circular orbits, where the tide is steady and cannot change the energy
of a star.

Such work can give a first answer to the question whether it is tidal
shocking or two-body relaxation which dominates the mass loss of
galactic clusters.  One answer, based on the work of Gnedin \&
Ostriker (1997), is given in Fig.5.  When we come to study simulations
in the next sub-subsection, this is a question on which we shall
concentrate. 

\begin{figure}
\plotfiddle{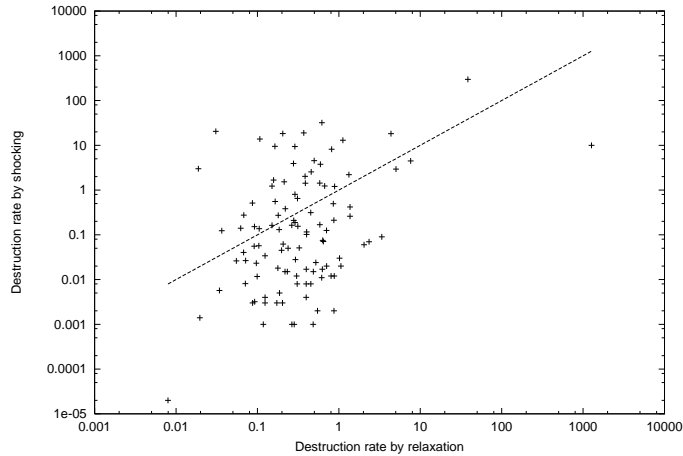}{2.5 in}{-90}{35}{35}{-150}{190}
\caption{Time scale of rate of destruction by shocking (bulge {\sl and disk}) against the
destruction time scale for relaxation, from  Gnedin \& Ostriker
(1997).  Each point represents one Galactic globular cluster from
their sample, using the ``OC Isotropic'' data in their
Table 3.  Units are per Hubble Time, and the line corresponds to
equality of the two destruction rates.  Note that both the most rapidly
and the least rapidly 
dissolving cluster are relaxation-dominated (below the line).}
\end{figure}

\subsubsection{Simulations of Bulge Shocking}  Much can be learned
from the motion of test particles in a fixed cluster-like potential
orbiting in a galaxy-like potential.  Keenan \& Innanen (1975) used a
King-like cluster model and both point-mass and axisymmetric galaxy
potentials, while Keenan (1981ab) used point masses for both.  This
case is the ``elliptic restricted three-body problem'', which has
experienced a modest renaissance recently in other contexts
(e.g. Benest 1998 and references therein).

Relaxation can be added to this simple model by incorporating
diffusion (Oh, Lin \& Aarseth 1992), and in this way Oh \& Lin (1992)
concluded that the presence of a tidal field could suppress escape.
Perhaps this result, which at face value contradicts all other wisdom
on this question, may be explained by the increase in the
escape time scale in a tidal field, as already discussed.

{\sl Self-consistent} calculations of tidal effects on clusters on
elliptical orbits were carried out by Chernoff \& Weinberg (1991),
though the initial models considerably exceeded their tidal radii, and
therefore perhaps relate better to young clusters than to the old
Galactic globular clusters.  Much the same may be said of the later
studies by Johnston, Sigurdsson and Hernquist (1999), though they took care to
quantify the effects of relaxation.  This was also done in the
simulations by Combes et al. (1999), though the choice of initial
conditions is too patchy to arrive at general conclusions on the
relative importance of bulge heating and relaxation.

An independent attack on this question was begun by Baumgardt (1997),
who computed some small $N$-body models of clusters on elliptical
orbits about a point mass galaxy with initial conditions similar to
those of Murali \& Weinberg (1997), i.e. the cluster is started at
{\sl apocentre} with a limiting radius equal to the tidal radius at
{\sl pericentre}.  

One attraction of these initial conditions is that they correspond to
one of the conclusions of Oh \& Lin (1992), which was that the
limiting radius of clusters is determined at pericentre and maintained
all along the orbit.  Another attraction is that they allow a
systematic exploration of the relative importance of tides and
relaxation.  Collaborative work with H. Baumgardt is currently under
way at Edinburgh on this problem (Fig.6).

One of the interesting questions which even this research will leave
unanswered is the behaviour of clusters in axisymmetric potentials.
The essential difference is that the perigalactic distance varies
considerably from one bulge passage to the next.  It may be that mass
loss through bulge shocking is much less regular and more infrequent
than is suggested by Fig.6.

\section{Summary and Conclusions}

Observation of tidal tails of Galactic globular clusters now seem well
established and consistent, but the detailed modelling and
understanding of various individual objects remains almost unexplored.
Kinematic data also suggests that tidal effects may impose an
observable signature, while it is suggested in this paper that stars apparently
observed above the escape energy may be genuine cluster members
trapped for long periods in the combined field of the cluster and
galaxy.  It seems certain that mass loss has altered both the stellar
mass function of each cluster and the distribution of cluster
properties throughout the galaxy.  The expected correlations among the
stellar mass functions are a little elusive, however, and it is only
in extreme cases that the effects of mass loss seem clear cut.

\begin{figure}
\plotfiddle{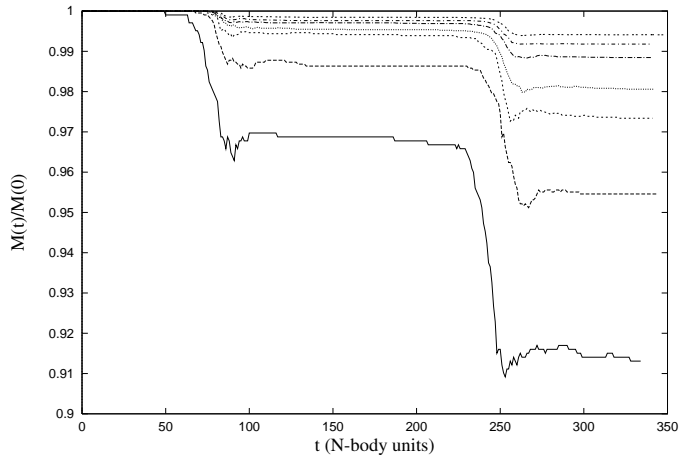}{2.5 in}{-90}{35}{35}{-150}{190}
\caption{Mass loss from a cluster on an elliptical orbit ($e=0.5$) about a
point-mass galaxy.  The initial model is a King model with scaled
central potential $W_0 = 3$, and other initial conditions are given in
the text.  $N = 1$K to $64$K in steps of a factor of $2$, from bottom
to top.  The potential is softened and all stars have equal mass.  For
low $N$ the rate of mass loss is approximately proportional to $1/N$,
as would be expected for relaxation-dominated mass loss; for large $N$
the rate appears to level off, as would be expected for mass loss
dominated by bulge shocking.  The two processes appear to be roughly
comparable for the largest $N$ here.
}
\end{figure}

There are still loose ends to be tied up in the theory of escape from
a cluster with a tidal cutoff:  is the escape time scale proportional
to the relaxation time, with the Coulomb logarithm, or is it dominated
by single energetic scatterings, as in the formula of H\'enon?  The
``induced'' loss of mass may help to understand the distinction
between ``skiing'' and ``jumping'' models, but a quantitative
understanding of this distinction is still missing.  When the tide is
represented by a proper external field, the effect of the large
population of ``potential escapers'' complicates the time scale for
mass loss, and a theory of this problem has recently been developed by
Baumgardt.  In the more realistic case of an elliptic galactic orbit,
an interesting focus for current research is the question of which
mass loss process dominates:  relaxation, or bulge shocking?  An
interesting future development will be to understand the effect of the
varying perigalactic distance in axisymmetric potentials.

\acknowledgments

I thank G.A. Drukier, G. Meylan and R. Wielen for additional
information and comment on their results, and H. Baumgardt for many
discussions on theoretical matters.  $N$-body calculations were
performed on hardware funded under SERC/PPARC grants GR/H93941 and
GR/J79461, and supplied by the GRAPE team at the University of Tokyo,
and partly using NBODY4 by S.J. Aarseth.  Work with H. Baumgardt is
supported by PPARC under grant 1998/00044.  I also thank the
organisers of Star2000 for their financial support.

\end{document}